\documentclass[nonacm, sigconf, screen]{acmart}
\renewcommand\footnotetextcopyrightpermission[1]{}
\usepackage{pifont}
\usepackage{graphicx}
\AtBeginDocument{%
  \providecommand\BibTeX{{%
    \normalfont B\kern-0.5em{\scshape i\kern-0.25em b}\kern-0.8em\TeX}}}

\usepackage{subcaption}
\usepackage{booktabs}
\usepackage{multirow}
\usepackage{tabularx}
\usepackage{float}
\usepackage{graphicx}
\usepackage{dblfloatfix}
\usepackage{footnote}
\usepackage{caption}
\usepackage{subcaption}
\usepackage[flushleft]{threeparttable}
\usepackage{soul}
\usepackage{newtxtext,newtxmath}

\setcopyright{none}
\copyrightyear{2023}
\acmYear{2023}
\acmDOI{--}

\begin{document}

\title{Self Supervised Vision for Climate Downscaling}

\author{Karandeep Singh}
\email{ksingh@ibs.re.kr}
\affiliation{%
  \institution{Data Science Group, IBS}
  \country{}
}

\author{Chaeyoon Jeong}
\email{lily9991@kaist.ac.kr}
\affiliation{%
  \institution{School of Computing, KAIST}
  \country{}
  }

\author{Naufal Shidqi}
 \email{naufal.shidqi@kaist.ac.kr}
\affiliation{%
 \institution{School of Computing, KAIST}
 \country{}
 }

\author{Sungwon Park}
  \email{psw0416@kaist.ac.kr}
\affiliation{%
  \institution{School of Computing, KAIST}
  \country{}
  }

\author{Arjun Nellikkattil}
\email{arjunbabun@pusan.ac.kr}
\affiliation{%
  \institution{{Department of Climate System, PNU}
  \country{}
  }}

\author{Elke Zeller}
\email{elkezeller@pusan.ac.kr}
\affiliation{%
  \institution{Department of Climate System, PNU}
 \country{}
  }

\author{Meeyoung Cha}
\email{mcha@ibs.re.kr}
\affiliation{%
  \institution{IBS \& School of Computing, KAIST}
  \country{}
  }

\renewcommand{\shortauthors}{Singh et al.}

\begin{abstract}

Climate change is one of the most critical challenges that our planet is facing today. Rising global temperatures are already bringing noticeable changes to Earth's weather and climate patterns with an increased frequency of unpredictable and extreme weather events.
Future projections for climate change research are based on Earth System Models (ESMs), the computer models that simulate the Earth's climate system.
ESMs provide a framework to integrate various physical systems, but their output is bound by the enormous computational resources required for running and archiving higher-resolution simulations. 
For a given resource budget, the ESMs are generally run on a coarser grid, followed by a computationally lighter ``downscaling'' process to obtain a finer-resolution output.
In this work, 
we present a deep-learning model for downscaling ESM simulation data that does not require high-resolution ground truth data for model optimization. This is realized by 
leveraging salient data distribution patterns and the
hidden dependencies between weather variables 
for an \textit{individual} data point at \textit{runtime}. 
Extensive evaluation with 2x, 3x, and 4x scaling factors demonstrates that the proposed model consistently obtains superior performance over that of various baselines. 
The improved downscaling performance and no dependence on high-resolution ground truth data make the proposed method a valuable tool for climate research and mark it as a promising direction for future research.
\end{abstract}

\keywords{Climate Downscaling, Super Resolution, Earth System Model, Orography, Topoclimatology, Topoclimatic Attention}


\settopmatter{printfolios=true}
\maketitle

\section{Introduction}

The increase in global temperatures is resulting in unprecedented changes in Earth's weather and climate patterns. Noticeable changes include an increase in the 
frequency of extreme weather events 
~\cite{ipcc_ex}, while less conspicuous changes include melting ice caps, rising sea levels, lowering crop yields, and impacts on oceans and the fishing industry \cite{ipcc_imp}. This phenomenon, better known as climate change, has critical consequences for Earth's ecology and the future of human societies.

Future climate projections are largely driven by Earth System Models (ESMs), the computer models that simulate the Earth's climate system. ESMs are a fundamentally important tool for climate research and the design of future climate policies and strategies\footnote{It is worth noting that weather projections are average weather conditions over a relatively short period (hours, days, a few weeks), whereas climate projections are average weather projections over a longer period (decades, centuries)}. High-resolution (HR) climate simulations provide detailed information at a regional level that is better suited for policy design at the local level. However, ESMs are constrained by the enormous computational resources required by the HR simulations. ESMs are generally run at coarser resolutions, and finer, regional-level information is derived by ``downscaling.''\footnote{We use the term downscaling to refer to the realization of finer-resolution output from a coarser simulation. We also limit our paper's scope to statistical downscaling.}

Climate downscaling (CD) is similar to super-resolution (SR) tasks in computer vision. While CD improves the resolution of climate data, SR improves the resolution of digital images. Deep learning and computer vision techniques have enabled major improvements in the SR task. Moreover, deep learning is an effective CD technique~\cite{Vandal_2017, Park_2022}. SR models are usually developed assuming that either paired or unpaired high-resolution (HR) and low-resolution (LR) data are available. Ground-truth HR data are also assumed to be accessible in \textit{sufficient} quantity so that a desired level of fit and generalizability of a model can be achieved. Furthermore, supervised models assume that the distribution of the test data is similar to that of the training data. These assumptions also hold for machine learning-driven downscaling, where the majority of proposed models are either fully- or partially-supervised~\cite{groenke2020climalign, Vandal_2017, Park_2022}. However, access to the data required for SR learning is not guaranteed in real-world climate research.

Moreover, generating LR-HR paired climate data is particularly challenging due to the inherent stochasticity in climate simulations. Minor changes in simulation runs can lead to vastly different outcomes, a phenomenon that is also known as the \textit{butterfly effect}~\cite{palmer2014real}.
Climate simulations can span thousands or even millions of years. It is computationally infeasible to generate a sufficient amount of HR simulation data for a supervised setting. For instance, an HR simulation with the Community Earth System Model (CESM) uses 250K core hours per simulated year, generating approximately one terabyte of data per compute day \cite{small2014new}. Another characteristic of climate data is the vastly different nature of climate variables. Thus, a model trained with limited HR data can lead to subpar performance and poor generalizability due to domain shifts.

This paper introduces a deep learning-based climate downscaling model that does not require HR ground truth data for training and adapts to the input data distribution at runtime.
The proposed method uses self-supervised learning to train an instance-specific model on a single data instance at runtime. We introduce three climate components in the model that adapt to the characteristics of the underlying weather variables and improve the transferability. 
  
The first component is \textbf{self-supervised pre-training}, which enables the transfer of knowledge across the dataset and reduces the overall run time. The second component is \textbf{channel segregation}, which enhances the learning of complex features specific to different weather variables that could be vastly different natures (as shown in Figure~\ref{fig:demo_fig}. The third component is \textbf{topoclimatic attention}, which forces the model to learn intricate inter-channel dependencies.
The latter two components help extract improved features from the climate simulation data and the physical relationship of climate variables.
The model is evaluated with multiple scaling factors by downscaling the temperature (TS) and total precipitation (PRECT) data obtained from CESM (v. 1.2.2) and utilizing the gradient of topography (dPHIS) as additional data.

Without reliance on ground truth HR data, our downscaling method has great potential for climate downscaling research. Immediate benefits include: (1) enabling larger ensemble runs to aid in simulation stability, 
(2) downscaling historic and temporally large simulations, (3) democratizing climate research and allowing organizations with limited access to computational power to run high-resolution simulation models\footnote{Model code and data will be made freely available: \url{https://github.com/k-s-b/climate_sd}}, (4) enabling ``temporal downscaling'' by running coarser models with higher temporal resolution, and (5) promoting a greener climate with a lower energy budget.  The main contributions of this paper are as follows:
\begin{itemize}

    \item Our climate downscaling model does not depend on high-resolution climate data, and outperforms baselines for the 2x through 4x downscaling factors.

    \item Our model is agnostic to underlying climate conditions and dynamically adapts to input data distribution at runtime.

    \item It is computationally lighter than physical models with the same resolution for climate downscaling.

    \item We propose climate data-specific adaptations to learn climatic patterns tied to topography.

    \item By the self-supervised pre-training method, our model considers climate knowledge transferability by transferring knowledge from other low-resolution data points while enabling much faster inference times.
\end{itemize}

\section{Related Work}

\subsection{Single Image Super-Resolution}
 
Image super-resolution (SR) task aims to enhance the resolution of a low-resolution (LR) image. Since the first introduction of a CNN-based SR network by Dong \textit{et al.}, CNN-based SR methods have led to impressive results~\cite{SRCNN2014Dong, Ledig_2017_CVPR, veryDeep2016, Lim_2017_CVPR_Workshops, haris2018deep,zhang2018residual}. These methods involve deep-CNN models that learn to directly map the LR images to high-resolution (HR) images. On the other hand, generative-based models, such as generative adversarial networks (GAN) ~\cite{Ledig_2017_CVPR}, and recent works based on diffusion models ~\cite{saharia2022image,li2022srdiff} are also utilized for this task, producing even more realistic HR images.

\subsection{Unsupervised Single Image Super-Resolution}
The methods mentioned above are fully supervised learning methods. 
Other branches of work, including unpaired SR or blind SR, extending the problem to situations where there are no direct LR-HR pairs, \textit{i.e.}, absence of ground truth~\cite{wang2021unsupervised, kernelestimation2020, Cornillre2019BlindIS, Zhang_2018_CVPR}.

However, those methods still rely on the availability of the HR image dataset. To deal with the lack of HR images, Zero-Shot Super-Resolution (ZSSR) takes a unique approach by using test images themselves as training images~\cite{ZSSR}. This approach degrades test images to generate synthetic LR and HR pairs and learns the mapping through these pairs.

\begin{figure}[t!]
    \centering
    \hspace*{-3mm}
    \includegraphics[width=1.07\linewidth]{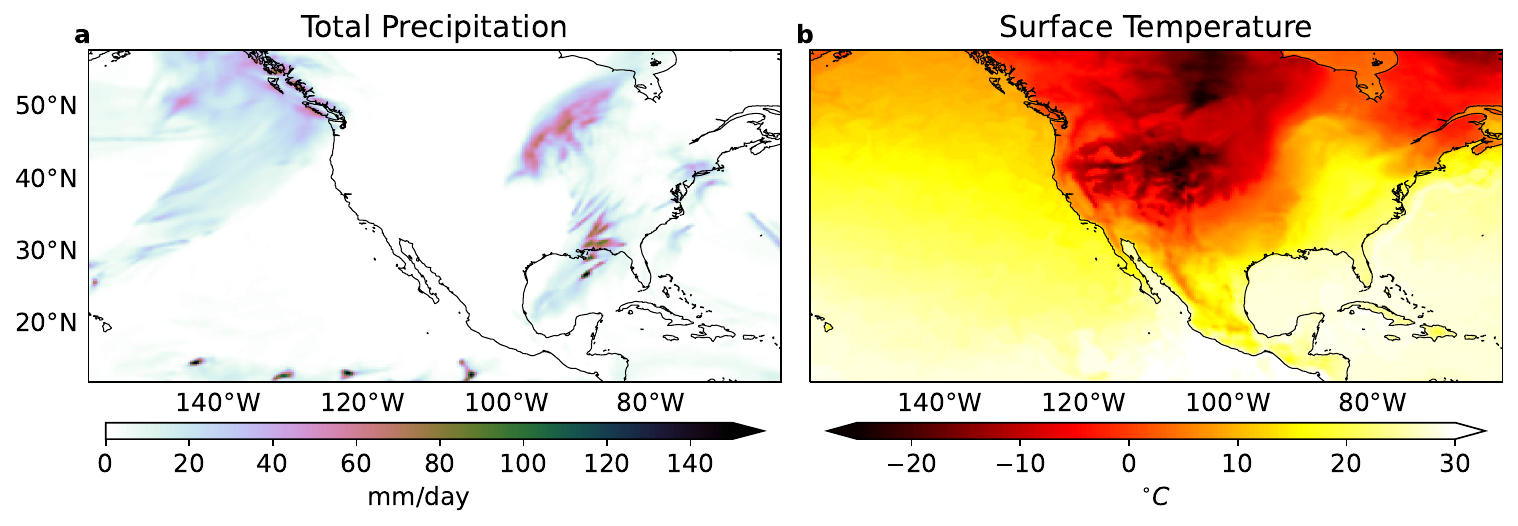}
    \caption{A snapshot of a time step representing total precipitation (PRECT) and surface temperature (TS). Both variables exhibit remarkably different behaviors.}
    \label{fig:demo_fig}
    \label{demo_fig}
\end{figure}

\subsection{Climate Downscaling using Deep Learning}
The use of deep learning for climate downscaling has been studied in previous research, including the DeepSD method, which uses a CNN architecture to downscale elevation and precipitation information~\cite{Vandal_2017}. Studies have also used SRCNN and other super-resolution (SR) methods to downscale climate features such as temperature, precipitation, ocean surface temperature, or rainfall~\cite{Ducournau2016, kumar2021deep,bano2020configuration}. Other works have used deep learning models like autoencoders, adversarial networks, normalizing flows, and LSTM networks to achieve climate downscaling~\cite{vandal2019intercomparison, stengel2020adversarial, bittner2023an, winkler2023climate, misra2018statistical}. Results presented in these works demonstrate that deep learning approaches can outperform conventional climate downscaling techniques.

However, applying SR methods from computer vision directly to climate data poses additional challenges since climate data has a different nature from regular images. Several recent studies have attempted to incorporate climate-specific attributes or physics into deep learning models, with promising results. The study of Chou \textit{et al.} uses temporal information to improve the resolution of monthly precipitation data by using LSTM-based network~\cite{Chou2021}. Park \textit{et al.} downscales precipitation and temperature information by SRResnet-based method with geospatial encoding and topographical information~\cite{Park_2022}. Rampal \textit{et al.} reports downscaling rainfall by using CNN with various combinations of input climate variables and numerous training losses to find which setting works the best for rainfall downscaling ~\cite{rampal2022high}. Harder \textit{et al.} utilize a hard constraining layer to mimic physical constraints in the model to downscale satellite data and standard weather datasets ~\cite{harder2022generating}.

\section{Problem Setting}

Statistical climate downscaling projects coarse-grained climate data at a finer, regional scale. This is accomplished by establishing a statistical relationship between LR--HR data pairs or generally by nonparametric approaches such as interpolation. Because HR data representative of climate patterns across the entire simulation are rarely available, we formulate the problem of climate downscaling in the absence of ground truth HR data as follows:\\

\noindent \textbf{Problem}: \emph{Given an LR climate data instance ${X}$, and no availability of LR-HR pairs, the goal is to obtain an HR version ${X}$$\uparrow$ of ${X}$ by climate downscaling such that the performance in terms of the root mean squared error (RMSE) and visual quality of downscaled data is improved.}\\

Climate simulation data are gridded, where the height and width of the input data represent the number of grid points over a spatial area. Higher grid points for an area represent higher resolution and better represent regional climatic conditions.
If an LR data instance for a given area has dimensions of $\mathbb{R}^{C \times W' \times H'}$, we downscale it to $\mathbb{R}^{C \times W \times H}$ such that $W = s \times W'$ and $H' = s \times H'$, where $s$ is the scaling factor ($s>1$), and $C$ denotes the number of weather variables in the data. 
In this work, $C=3$ with temperature (TS), total precipitation (PRECT), and topography gradient (dPHIS) as the weather variables. 
While temperature and topographic height influence the precipitation characteristics, using the elevation directly tends to introduce opposing precipitation biases on the windward and leeward sides of mountains \cite{Park_2022}. This has led us to incorporate topographic information in the form of the topographic height gradient which helps reduce these biases. Our approach is agnostic to the type and number of input weather variables and can downscale multiple variables at once.

\subsection{Data}
Real LR or HR data can be defined as data acquired from a process that \textit{originally} resulted in that quality. 
Examples of real LR vs. HR include coarse- vs. fine-grained simulations, blurry vs. crisp images captured by a shaky vs. stable camera, and images acquired by a low- vs. high-resolution camera.
Because obtaining real climate LR-HR pair data is particularly challenging, synthetic LR data are often obtained by degrading real HR data. This also makes it possible to perform a quantitative analysis of the model performance.

The climate simulation data in this work are sourced from HR simulations by the Community Earth System Model (CESM) version 1.2.2~\cite{hurrell2013community, small2014new}. The CESM is a fully coupled global climate model that is primarily maintained by the National Center for Atmospheric Research (NCAR). The data represent the daily means of present-day climate conditions for 140 years and have a spatial resolution of approximately 0.25$^{\circ}$ in the atmosphere component and 0.10$^{\circ}$ in the ocean components~\cite{chu2020reduced}. Only the last 20 (7300 data points) years of data are used for this work so that the simulation model reaches an equilibrium state. Surface temperature (TS), total precipitation (PRECT), and topography gradient (dPHIS) are the three climate variables selected from these data and used as the three-channel input. Only the TS and PRECT channels are downscaled, as the dPHIS does not change during the simulation time. Surface temperature and total precipitation are critical variables that determine weather and climate conditions and impact climate adaptation and mitigation policies. The TS and PRECT channels are depicted in Figure~\ref{fig:demo_fig}.

\begin{figure}[t!]
    \centering
    \hspace*{-3mm}
    \includegraphics[width=1.1\linewidth]{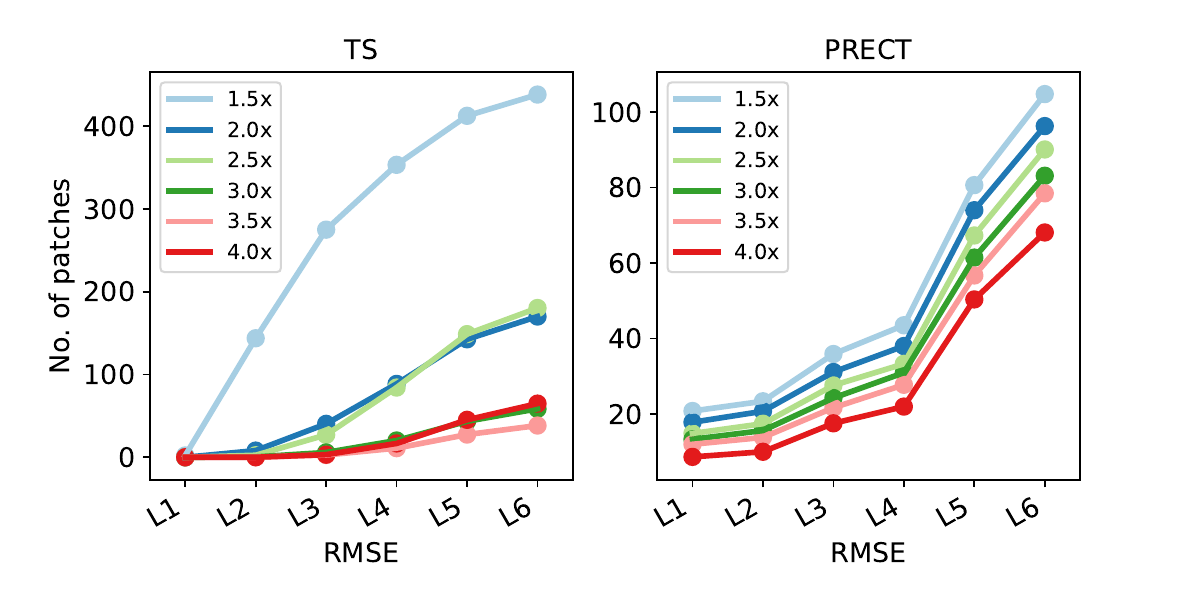}
    \caption{Information repetition across scales in climate data. The similarity between the 10 x 10 patches in the source image is measured across multiple scales (1.5x--4.0x). y-axis represents levels L1 to L6 - the increasing RMSE values for the cutoff threshold that determines the level of similarity. The cutoff range is 1E-2 to 1E-1 for TS and 0.5E-14 to 1E-12 for PRECT. the x-axis represents the number of patches that have RMSE values lower than the cutoff value at that level.
    }
    \label{fig:ts_prect_patches}
    \label{fig:prectpatches}
\end{figure}

\subsection{Information Redundancy in Climate Data}
\label{sec:info_repeat}

Natural images have been shown to possess recurring patches of information across their original images as well as their coarser versions~\cite{glasner2009super}. The internal information of an image is known to have stronger predictive power than the information from external databases~\cite{zontak2011internal}. 
Inspired by previous super-resolution works such as~\cite{ZSSR}, that leveraged the internal information of a natural image, we analyze information redundancy in the climate data.

Downscaling algorithms in climate research should process and make predictions based on data in the standard units of target weather variables (such as Kelvin for temperature, and cm/day for precipitation). Therefore, our model processes the raw climate data. The data recurrence analysis is also performed with raw data. Figure~\ref{fig:prectpatches} presents the statistics of this analysis for $10 \times 10$ patches across multiple scales for $100$ randomly chosen instances from our dataset. It can be seen that climate data have information redundancy for both the TS and PRECT channels. This behavior can be understood as a repetition of weather patterns across different resolutions of climate data, which can be leveraged for the SR task.

\begin{figure*}[t!]

\centerline{\includegraphics[width=1.0\linewidth, trim = 1cm 5.5cm 2cm 1cm, clip]{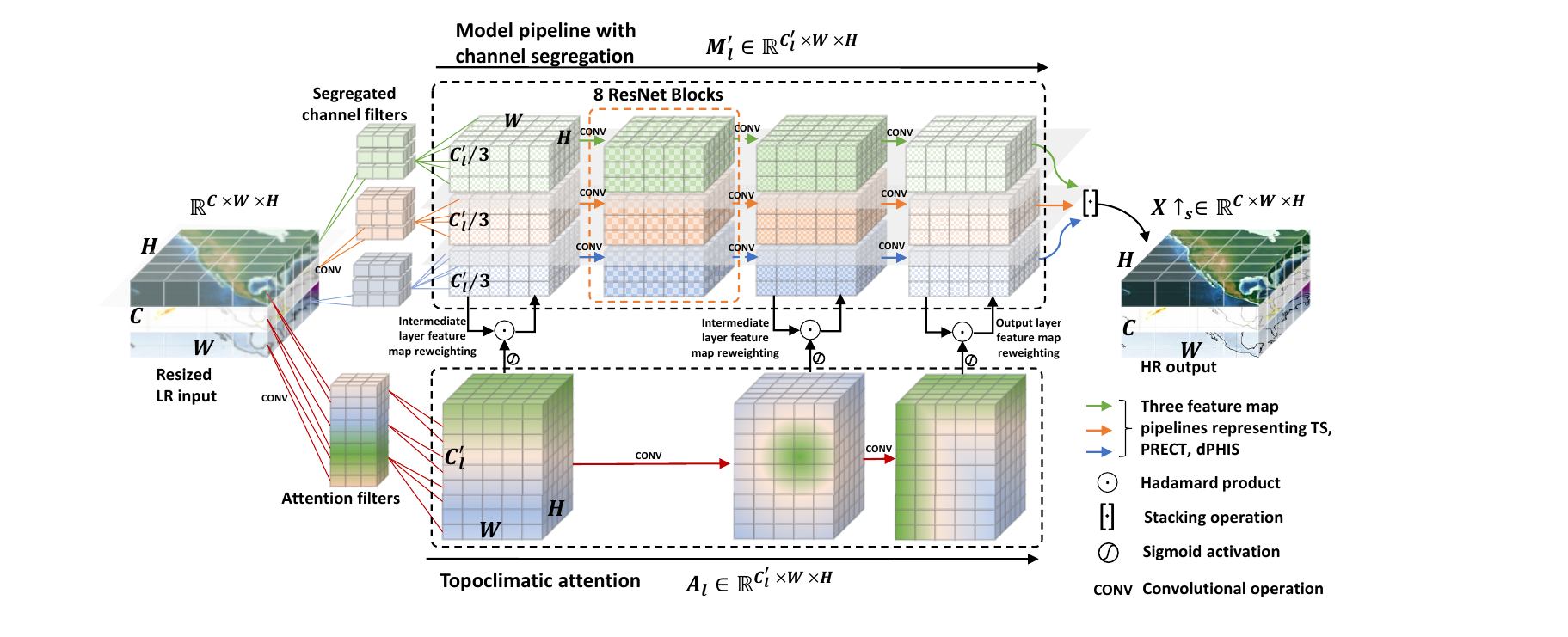}}
\vspace{-5mm}
\caption{Core architecture of the proposed model. A low-resolution input ${X}$ is downscaled to ${X}$$\uparrow_{s}$ ($s>1$) and (${X}$, ${X}$)$\uparrow_{s}$ $\in  \mathbb{R}^{C \times W \times H}$ where $C=3$, represent three weather variables: temperature (TS), total precipitation (PRECT), and gradient of topography (dPHIS). ${X}$ is resized to target size before feeding into the model.
The upper half of the image shows the channel segregation with three pipelines, with respective filters for each input variable. The topoclimatic attention mechanism is the lower half parallel to the main model flow. The attention mechanism re-weights the intermediate feature maps at multiple stages and enforces learning of topoclimatic relationships.}
\label{fig:architecture}
\end{figure*}

\section{Model}
 
The proposed model is a self-supervised CNN model that operates without the need for the ground truth dataset. The core idea is to leverage internal information of the climate data while using channel segregation and topoclimatic attention to enhance feature learning for weather variables. Figure~\ref{fig:architecture} depicts the core architecture of the model.

Given the LR input $X$, we first generate pseudo-LR data $X\downarrow_{d}$ by blurring $X$. A deep learning model is then trained on the generated data pair (Section \ref{sec:training_process}). 
We use self-supervised pre-training to aid knowledge transfer from the entire input LR data and propose a new network architecture that is tailored for adapting salient characteristics of climate data (Section \ref{sec:topoclimate}).

\label{sec:ssl}
\subsection{Training Process}
\label{sec:training_process}
Section~\ref{sec:info_repeat} shows the information repetition across multiple scales of climate data, indicating what an HR version of an LR patch might look like. This forms the basis for example--based SR across different scales of a data point~\cite{glasner2009super}, where a CNN model can be trained to learn a generalized mapping from a downgraded version of the LR input back to its original resolution. Specifically, if ${X}$ is the LR input data point, and ${X}$$\downarrow_{d}$ its downgraded version by a factor $d$ $(d<1)$, an instance--specific CNN model $f_\theta$ is then trained on (${X}$$\downarrow_{d}, {X}$) pair to obtain ${X}$$\uparrow_{s}$ by applying the trained model on ${X}$, \textit{i.e.}, ${X}$$\uparrow_{s} = f_\theta(X)$. Here, $s$ indicates the target downscaling factor and $d$ is the downgrading factor, with $d=$ $1 \over s$ typically. Training a data point-specific model enables it to adapt to data distribution patterns salient to that data point. It also makes our approach agnostic to different variables and means--states of climate. \smallskip
 
\noindent{\textbf{Self-supervised training using pseudo data pair:}} 
\label{sec:pseudo_data_pair}
A downscaling model ideally should learn the mapping between low-resolution (LR) and high-resolution (HR) data pairs, but we assume HR data is not accessible. Similar to~\cite{ZSSR}, we generate the ``pseudo'' LR--HR pair by degrading the original high-resolution data (${X}^{HR}$). We regard this degraded data as the low-resolution model input ($X$) to be downscaled and apply degradation to generate ${X}$$\downarrow_{d}$. The downgraded versions can be prepared by a desired kernel such as blurring the data with bicubic interpolation. The data pair ${X}$$\downarrow_{d}$ -- ${X}$ now becomes a pseudo LR (${X}$$\downarrow_{d}$) and pseudo HR ($X$) pair and is used as training data. These data generation steps are summarized below:

\begin{enumerate}
    \item \textsf{Orginal HR (${X}^{HR}$)}: The orginal HR data from CESM (0.25$^{\circ}$ resolution) 
    \item \textsf{Model input or pseudo HR (${X}$)}: Downgradation of ${X}^{HR}$ by factor $d$
    \item \textsf{Pseudo LR (${X}$$\downarrow_{d}$)}: Downgradation of ${X}$ by factor $d$
    \item \textsf{Downscaled model output (${X}$$\uparrow_{s}$)}: Downscaling by factor $s$ by model inference over ${X}$ 
\end{enumerate}
Before passing as an input to the model, ${X}$ and ${X}$$\downarrow_{d}$ are re-scaled to the target downscaling size. i.e., the dimension of the input and the output data is preserved. The optimization objective of the model becomes minimizing:
\begin{equation}
\mathcal{L} = \dfrac{1}{C \times W \times H} \sum_{C,W,H} \mathcal{P}(f_\theta(X\downarrow_{d}), {X})
\label{eq:zssr_loss}
\end{equation}
where $\mathcal{P}$ is element-wise loss such as mean squared error (MSE) or mean absolute error (MAE).
Data can be augmented by choosing different values of $d$ and $s$ for steps 
2, 3, and 4, in addition to the target downscaling factor $s$ and $d$=$1/s$.

\subsection{Self-supervised pre-training}
\label{sec:mzsr}

Training a model with a single data instance carries some drawbacks: First, the convergence time can span long runs, making it infeasible to downscale a large number of data points in a reasonable time. Instead of starting from \textit{scratch}, adapting a model from the prior knowledge can greatly reduce the runtime.
Second, although an instance-specific model adapts well to the input, it misses the utilization of knowledge from other data points that may help improve the performance.

A deep learning model trained on a bigger dataset for one task can be adapted to various similar downstream tasks, which is also known as transfer learning.~\cite{finn2017model, sun2019meta, soh2020meta}. Inspired by this concept, we introduce a self-supervised pre-training step whereby we train a single model for the entire LR dataset by learning a mapping between the downgraded version of the LR data, (all ${X}$$\downarrow_{d}$ -- ${X}$ pairs; i.e., a low-resolution version of the entire LR data) and the LR data itself. The main difference between the instance-specific self-supervised model and this step is that the former is a single model optimized for a single data instance, while the latter is a single model optimized for the entire dataset.

Further, the pre-trained model is finetuned on randomly sampled data points that are augmented by random uniform noise. 
This step not only acts as a regularizer but also optimizes the pre-trained model for diverse scenarios. 
Both these steps aim to enable knowledge transfer from and across the input data, while greatly increasing the training efficiency. 

The downscaling task is realized by loading this pre-trained model and optimizing it for a single data instance on the go (individual ${X}$$\downarrow_{d}$ -- ${X}$ pairs). The effect of self-supervised pre-training on performance and runtime is detailed in the results section~\ref{sec:results}.

\subsection{Channel Segregation}
\label{sec:backbone}

Figure~\ref{demo_fig} presents a snapshot of the TS and PRECT channels and highlights a striking difference between the features of the two channels. The TS channel has a smoother gradient, gradually decreasing values from the equator toward the poles and spanning the entire spatial extent. In contrast, PRECT is non-linear and localized, and it has a high degree of variance within localized regions. 
This characteristic contrasts with natural images, where different channels, albeit with different values, represent measurements of the same snapshot. The standard convolutional neural network (CNN) models operate under this assumption since a weighting filter convolves over the entire input volume to output a unified representation as another volume.

Several issues arise while optimizing a CNN model with weather variables with heterogeneous features constituting the input volume. For instance, the model may be forced to simultaneously optimize for a low-frequency TS feature and a high-frequency PRECT feature at the same spatial location. This perplexity can lead to unwanted gradient flows during backpropagation, resulting in undesirable artifacts in model output (refer Figure~\ref{fig:artifacts})

However, the CNN should adapt to a particular weather variable's low- and high-level characteristics and leverage them for the super-resolution task. A mechanism is needed to handle the vastly diverse properties of each channel effectively.
We enable it by constraining our CNN model to learn separate weights for each input weather variable. Our model can be viewed as consisting of $C$ different pipelines (\textit{i.e.}, one for each input channel) running in parallel, which are stacked at the last layer and optimized simultaneously. If $C'$ represents the number of channels in the intermediate feature map, then each channel is convolved by a filter of size $C' \over C$ where $C' = I \times C$ and $I$ is an integer. This CNN operation, also known as depth-wise-convolution, can be represented as:
\begin{equation}
 {\mathbf{Out}(N,C')} =  \left[ { \sum_{c} \mathbf{Filter}(C',c) \cdot \mathbf{Input}(N,c) + \mathbf{b}_{out}} \right]_{\forall c\ \in\ C}
\label{eq}
\end{equation}
where $N$ is the batch size, $C$ the number of weather variables, $C'$ the number of output channels, and [$\cdot$] represents the stacking operation. This architecture can learn features for each channel by reducing the undesired artifacts in the model output.

\subsection{Topoclimatic Attention}
\label{sec:topoclimate}
Separating the filters for individual channels can improve performance but has the drawback of missing out on inter-channel dependencies. This is particularly true for weather data, where intricate relationships between temperature, precipitation, and topography can exist. To account for such relations, we propose a topoclimatic attention mechanism that learns the cross-channel dependencies and enforces them by reweighting low-, intermediate-, and high-level feature maps in the main model pipeline. We define low- to high-level feature maps as the output of the CNN model from the first through the last layers.
The attention mechanism learns weights $\mathbf{A}_{l}$ ($l \in L$, where $L$ are the model layers) that encompass these dependencies with the following steps. \smallskip

\noindent \textbf{Convolution over volume:} If $\mathbf{M'}_{l} \in  \mathbb{R}^{C_{l}^{'} \times W \times H}$ is the feature map at the layer $l$, then $\mathbf{A}_{l} \in  \mathbb{R}^{C_{l}^{'} \times W \times H}$ is the attention weight at layer $l$, computed by a convolution over the entire input volume. A separate set of weight filters are trained for computing $A_{l}$. Spanning the entire volume layer-wise enables learning of dependencies between the weather variables at a given level $l$ of the feature map. \smallskip

\noindent \textbf{Generation of attention weights:} At given level $l$ the attention weights $A_{l}$ can be represented as: 
\begin{equation}
 {\mathbf{A_{l}}} =   \sigma\Bigl(\phi\Bigl(\mathbf{Conv}^{l} \Bigl(\phi \Bigl( ... \Bigl(\phi (\mathbf{Conv}^{1}\Bigl(\mathbf{X}\Bigl)\Bigl)\Bigl)...\Bigl)\Bigl)\Bigl)\Bigl)
\label{eq}
\end{equation}
where $\mathbf{Conv}^{l}$ is a 2D covolutional operator at layer $l$, $\phi$ and $\sigma$ are LeakyReLU and Sigmoid activation functions respectively.
If $A'_{l}$ represents the raw output of the convolution operation at layer ${l}$ (\textit{i.e.} without the Sigmoid activation) then $A_{l}$ ($l > 1$) is computed by passing $A'_{l-1}$ through  $\mathbf{Conv}^{l}$. At $l=1$, the $A_{l}$ is computed over the input. To enable this operation to learn complex non-linear correlations, the convolved output is transformed by the LeakyReLU non-linear activation function. Finally, the entire output is normalized in the range [$0$, $1$] by passing it through the sigmoid activation function. \smallskip

\noindent \textbf{Feature map reweighting:} The intermediate feature map $ \mathbf{M'}_{l}$ is updated to $ \mathbf{M}_{l}$ by attention weight $A_{l}$ as:
\begin{equation}
 {\mathbf{M_{l}}} = \mathbf{A}_{l} \odot \mathbf{M'}_{l}
\label{eq}
\end{equation}
where $\odot$ is the Hadamard product, and $\mathbf{M'}_{l}, \mathbf{M}_{l}, \mathbf{A}_{l} \in \mathbb{R}^{C_{l}^{'} \times W \times H}$. The attention mechanism reweights the feature maps to highlight the importance of inter-weather variable dependencies and the regions of the highest importance.

\section{Experiments and Results}
\label{sec:results}
The architecture and components of the model are intended to provide optimal performance with a short inference time. The specifications of the implementation are outlined below. \smallskip


\noindent \textbf{Data}: The original HR data have a spatial resolution of $0.25^{\circ}$ ($25 \times 25$ km) and dimensions $213 \times 321 \times 3$. TS, PRECT, and dPHIS are the three channels, with $213 \times 321$ grids each. Data are degraded by bicubic interpolation and are normalized by calculating the mean and standard deviation for each $X$ individually. We use a scale factor of $2$ when generating pseudo-LR and HR pairs in the pre-training and finetuning phases. Different scale factors (2x, 3x, 4x) are used when fine-tuning the model on each data instance. \smallskip

\noindent \textbf{Model pipeline}: The number of feature map channels is set as $63$ with ResNet architecture. Because of channel segregation during convolution, the number of channels in the feature map should be a multiple of input channels. For each channel, there are $21$ ($63$/$3$) effective channels in the intermediate feature map. The first CNN layer is followed by $8$ ResNet blocks, another CNN layer, and an output layer. Each data point has its own fine-tuned model as described in Section~\ref{sec:mzsr}. \smallskip

\noindent \textbf{Self-supervised pre-training}: 
The training is performed in three steps. The pre-training training step is performed for 150 epochs at a learning rate of 1e-4. Next, the finetuning step on noisy samples is run at 1,000 epochs, with a learning rate of 1e-2. For the final inference, fine-tuning on a single data point is completed over 20 epochs, with a learning rate of 1e-4. \smallskip

\noindent \textbf{Attention}: We deploy feature reweighting with a topoclimatic attention map at the input layer, the layer proceeding residual block, and the output layer. Applying attention weights to the residual block increases the computational complexity without further enhancing the performance. Its CNN architecture is similar to that of the model pipeline. \smallskip

\noindent \textbf{Additional details}: The architecture is fully convolutional and conserves the input size by setting the kernel size to $3$ and stride and padding to $1$. Instance normalization is used for regularization. Adam optimizer and LeakyReLU with a negative slope of $0.2$ are used in the model and the attention pipeline. \smallskip

\noindent \textbf{Evaluation}: We repeat experiments over different scaling factors and compare the results with baselines. 
The model evaluation is performed with the \textit{entire} CESM dataset (7300 data points of size $213 \times 321 \times 3$), as the ground truth data is not exposed to the model at any point during the training. The results are reported over 2x, 3x, and 4x scaling factors, representing downscaling from resolutions of 50km, 75km, and 100km, respectively, to 25km. The model performance is measured by the root mean squared error (RMSE) between the downscaled model output and the target real HR data. 

\begin{table}[t!]
\centering
\begin{threeparttable}
\caption{Performance comparison against standard approaches and deep learning models for 2x, 3x, and 4x scaling factors for temperature (TS) and precipitation (PRECT, units are in 1e-8). The reported metric is RMSE (lower is better).}
\label{tab:result_comparison}
\setlength{\tabcolsep}{2.5pt}
\begin{tabular}{@{}c|cc|cc|cc@{}}

\toprule
\textbf{Scale}   & \multicolumn{2}{c|}{\textbf{2x}} & \multicolumn{2}{c|}{\textbf{3x}}  & \multicolumn{2}{c}{\textbf{4x}} \\ \midrule
\textbf{Channel} & \textbf{TS}     & \textbf{PRECT} & \textbf{TS}     & \textbf{PRECT}   & \textbf{TS}     & \textbf{PRECT}  \\ \midrule
Bilinear         & 0.6655 & 2.2302 & 1.1922 & 3.5510 & 0.8662 & 3.7215        \\
Bicubic          & 0.6829          & 2.0246  & 1.2491 &  3.6584         & 0.8647          & 3.5651        \\
PT*         & 0.4879          & 1.5046   & 0.9343 &  3.0128      & 0.7534 &  3.3845    \\
PT+FT*    & 0.4607          & \textbf{1.3552}   & 0.9109 &  2.8722      & 0.7435          & 3.4147    \\
Ours             & \textbf{0.3585} & 1.3795  & \textbf{0.7959} & \textbf{2.8703} & \textbf{0.7012} & \textbf{3.3837} \\ \bottomrule
\end{tabular}
\begin{tablenotes}
      \small
      \item *PT refers to inference followed by a self-supervised pre-training model only.
      PT+FT finetunes on individual data points but without channel segregation and attention components.
    \end{tablenotes}
\end{threeparttable}
\end{table}

\begin{table}[htp]
\centering
\caption{Performance comparison of the statistical method, supervised and self-supervised deep learning models based on 1000 data points.}
\label{tab:result_comparison_supervised}
\setlength{\tabcolsep}{2.2pt}
\begin{tabular}{@{}c|c|cc|cc@{}}
\toprule
   & \textbf{Scale}   & \multicolumn{2}{c|}{\textbf{2x}}  & \multicolumn{2}{c}{\textbf{4x}} \\ \cmidrule{2-6}
   &\textbf{Channel} & \textbf{TS}     & \textbf{PRECT}  & \textbf{TS}     & \textbf{PRECT}  \\ \midrule
 Interpolation & Bicubic          &  0.6804 & 2.0618 & 0.8616 & 3.6487       \\ \midrule
Supervised & GINE\textsubscript{5}      & 1.0712  & 1.7408 & 1.0859 & 3.5636      \\ 
   & GINE\textsubscript{10}         & 0.8112 & \textbf{1.2454} & 0.9388 & 3.5069     \\ \midrule
Self-supervised & Base*          & 0.5119          & 1.3367 & 0.7844          & 3.5329     \\
  & PT+{FT}          & 0.4580          & 1.3835          & 0.7399          & 3.4985     \\
 & Ours             & \textbf{0.3568} & 1.4607          & \textbf{0.6980} & \textbf{3.4749}\\ \bottomrule
\end{tabular}
\begin{tablenotes}
      \small
      \item *Base refers to an instance-specific self-supervised model such as~\cite{ZSSR}, with architecture as in Fig.~\ref{fig:architecture}, but without pre-training, channel segregation, or attention components. 
    \end{tablenotes}
\end{table}

\subsection{Performance Evaluation}
\label{sec:perf_eval}

Table~\ref{tab:result_comparison} shows the model performance over 2x, 3x, and 4x scaling factors for the entire dataset (7300 datapoints).
Linear and bicubic interpolation are commonly used methods in climate downscaling research. We also compare the model performance against that of strong baselines such as self-supervised pre-training (PT), which is a generalized model (without finetuning) with all ($X\downarrow_{d}$, $X$) pairs. Additionally, we evaluate our model against another baseline (PT+{FT}) where the pre-trained model is fine-tuned with individual data points but does not contain channel segregation and topoclimatic attention components. 

For the 2x scaling factor, our model outperforms all baselines, with a 47.50\% (TS) and 31.86\% (PRECT) reduction in the RMSE over that of bicubic interpolation and a reduction of 26.52\% (TS) and 8.32\% (PRECT) over a generalized self-supervised pre-training model (PT). 
The overall runtime is reduced greatly, as discussed in Section~\ref{sec:time_eval}. Except for the 2x scaling factor of PRECT, the full model consistently outperforms the baselines for the 3x and 4x scaling factors.

We finally compare our model performance with GINE\cite{Park_2022}, a supervised model for climate downscaling. GINE is trained with limited ground truth data in a supervised fashion by using 5\% (GINE\textsubscript{5}) and 10\% (GINE\textsubscript{10}) HR ground truth data ($X$-${X}^{HR}$ pairs). As GINE uses HR data for training, we compare its performance against 1000 randomly chosen data points from the held-out set.
The results in Table~\ref{tab:result_comparison_supervised} illustrate that our model outperforms the baseline, and even GINE trained with some HR ground truth data, although GINE performs better for PRECT when using 10\% labels.

The TS and PRECT channels do not always exhibit a similar tendency in terms of performance improvement. As previously mentioned, the two climate variables have distinct natures. The performance improvement in one channel leads to a minor performance trade-off in the other channel.
More about this in Section~\ref{sec:discussion_channels}.

\begin{table}[tp]
\centering
\begin{threeparttable}
\caption{\small Time efficiency of Base model, pretrained model without channel segregation and topoclimatic attention, and our model. The experiment is performed over 1000 data points and estimated for the entire 7300 data points and per each data point.}
\label{tab:efficiency_compare}
\setlength{\tabcolsep}{2.2pt}
\begin{tabular}{@{}c|c|c|c|c@{}}
\toprule
\textbf{Model}   & \textbf{Training}  & \textbf{Testing}       & \textbf{Full Testing} & \textbf{1 Data Testing}\\ 
\textbf{} & \textbf{Time (hr)} & \textbf{Time (hr)} & \textbf{Time* (hr)} & \textbf{Time* (sec)}  \\ \midrule
Base             & -                   & 38.8000              & 283.2394 & 139.68\\
PT+{FT}          & 3.1667                & 0.1693                & 1.2360 & 0.61\\
Ours            & 15.4167                 & 0.2821                & 2.0592  & 1.02              \\ \bottomrule
\end{tabular}
\begin{tablenotes}
      \small
      \item *Time is estimated from 1000 data points for actual testing time.
    \end{tablenotes}
\end{threeparttable}
\vspace{-3mm}
\end{table}

\subsection{Runtime}
\label{sec:time_eval}
As climate simulations usually cover a large number of timesteps, the downscaling method should have a low overall runtime for practical use.
We assess the overall computation times for a model trained on a single data instance from scratch (Base), a pre-trained and fine-tuned model without channel segregation and topoclimatic attention components (PT+FT), and a full model with these components (Ours). The Base model requires thousands of epochs for convergence. Due to the computational bottleneck induced by Base model, this particular assessment is only performed with randomly selected (seeded) 1000 data points\footnote{Please note that due to the computational bottleneck of the Base model, only the analysis in Tables~\ref{tab:result_comparison_supervised}, \ref{tab:efficiency_compare}, and \ref{tab:noise_robustness} is performed for same set of randomly chosen 1000 datapoints, whereas other results are presented for the full dataset.}. A single NVIDIA Tesla A100 GPU is used for running the models.

As evident in Table~\ref{tab:efficiency_compare}, the proposed methodology has a significantly reduced overall runtime, while also delivering improved RMSE values, as presented in Section~\ref{sec:perf_eval}. The inference time for 1 data point in the full model is 1.02 seconds, as compared to 139.68 seconds with the Base model, making the proposed methodology more practical as the number of downscaling data points increases.
The CESM 1.2.2 outputs approximately three model years per day with a state-of-the-art supercomputing framework. However, a climate model must be spun up for at least hundreds of years to obtain meaningful data~\cite{small2014new}.

\begin{table}[tp]
\centering
\caption{Ablation study on removing each architecture component (channel segregation, topoclimatic attention) and noisy fine-tuning process.}
\label{tab:ablation}
\resizebox{0.8\columnwidth}{!}{
\begin{tabular}{@{}l|cc@{}}
\toprule
\textbf{Scale}   & \multicolumn{2}{c}{\textbf{2x}}  \\ 
\textbf{Channel} & \textbf{TS}     & \textbf{PRECT}   \\ \midrule
Full model & \textbf{0.3585}        & 1.3795    \\
(-) Channel Segregation (Sec. \ref{sec:backbone})      & 0.4055   & \textbf{1.3245}       \\
(-) Attention (Sec. \ref{sec:topoclimate})         & 0.3656   & 1.4539 \\ \midrule
(-) FT & 0.4132 & 1.4106  \\\bottomrule
\end{tabular}
}
\end{table}

\subsection{Ablation Study}
\label{sec:ablation}

\begin{figure}[t]

\hspace*{-3mm}
\includegraphics[width=1.05\linewidth]{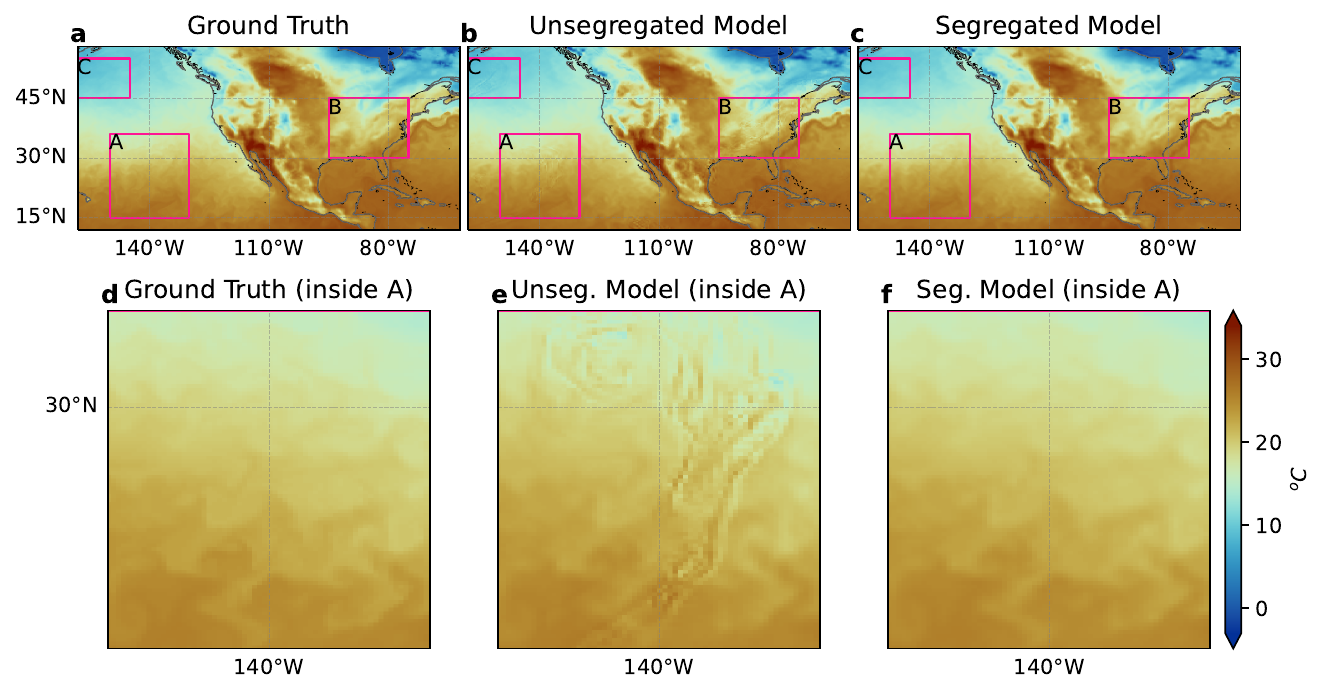}
\centering
\caption{ 
Ground truth high-resolution data (a), compared to downscaled output without channel segregation (b) downscaled output with channel segregation (c). Regions with noticeable artifacts are marked with pink rectangles. Region A is zoomed-in Figures (d), (e), and (f). RMSE values for (b) and (c) are smaller than bicubic interpolation, but model output without channel segregation suffers from undesirable artifacts.}
\label{fig:artifacts}
\end{figure}

The ablation study is performed by removing each of the proposed components while maintaining the same settings for the rest of the model. Specifically, we remove the channel segregation, topoclimatic attention, and finetuning on augmented data in the self-supervised pre-training and finetuning on individual data points (represented by (-) FT) and measure the model performance for all data points. As can be observed in Table~\ref{tab:ablation}, each of the model components is essential for the best performance. However, it can be seen that removing the channel segregation results in some performance gain for PRECT over that of the full model. At the same time, it also causes a performance drop in the TS channel. This implies that after an initial reduction in the RMSE, the TS and PRECT performance exhibit orthogonality to some extent. Nonetheless, the performance is better than that of the most commonly used baselines. Moreover, the quality of the downscaled output is substantially improved by our model, which can be seen in Figure \ref{fig:artifacts}.  Mitigating this trade-off will be an interesting future research direction.

\subsection{Model Robustness}
\label{sec:robustness}
We now explore the robustness of the model in various settings.

\noindent \textbf{Non-ideal blur kernel}:
We test the model's ability to downscale climate data when the blur kernel is non-ideal (\textit{e.g.}, not bicubic). To realize a non-ideal kernel, we inject \textit{freshly} generated uniform random noise into $X$ and ${X}$$\downarrow_{d}$, \textit{independently} for each data point. A noisy setting implies that the relationship between $X$ and ${X}$$\downarrow_{d}$ can be different from ${X}$$\downarrow_{d}$ and $X^{HR}$. This step is included in the self-supervised pre-training, and instance-wise finetuning phases as well.
The results are shown in Table~\ref{tab:noise_robustness}. Our model is robust against a non--ideal blur settings and still outperforms the baselines by a substantial margin.

\begin{figure}[t]
\includegraphics[width=0.65\linewidth]{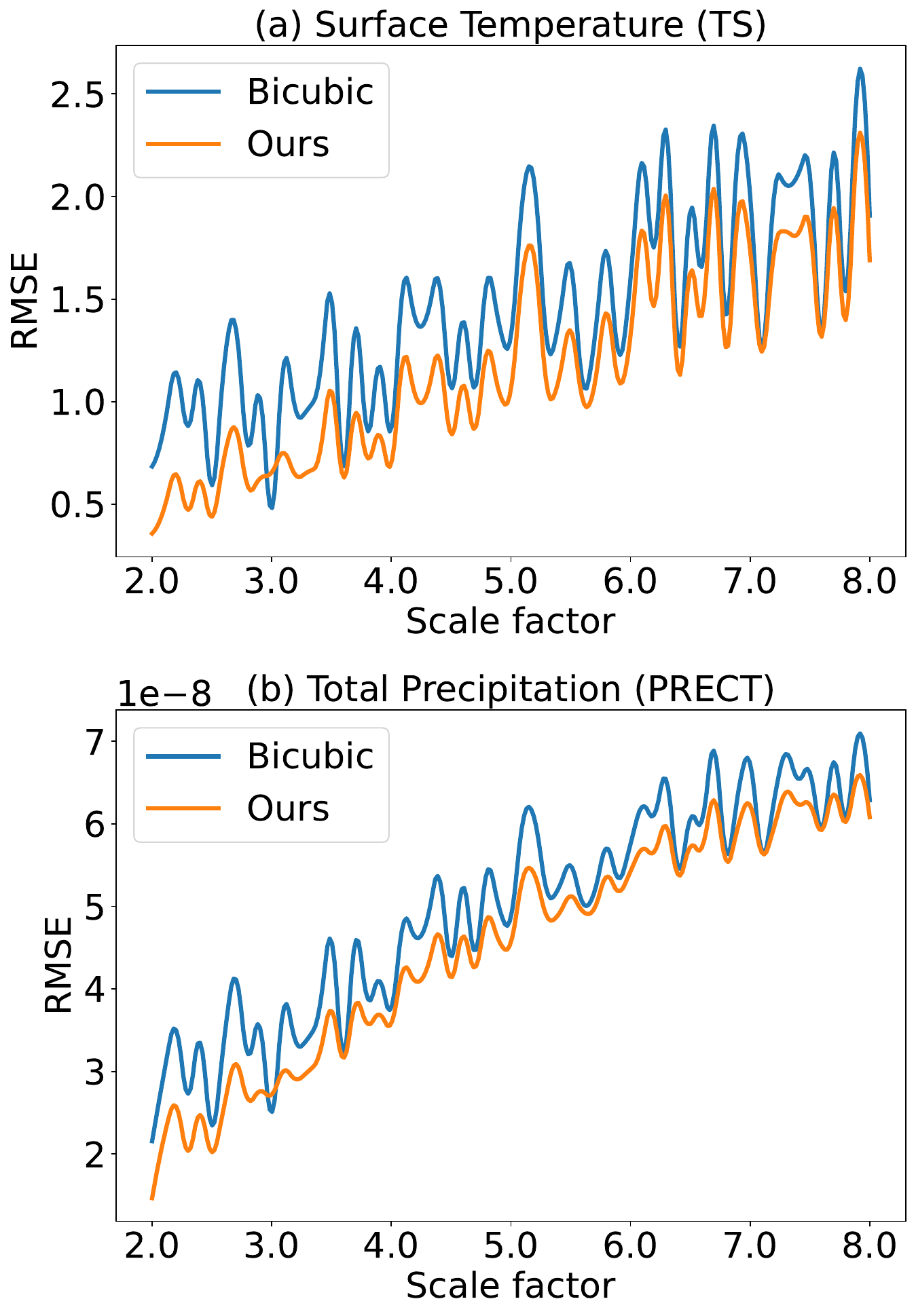}
\centering
\caption{Performance comparison for temperature (TS) Figure (a) and precipitation (PRECT) Figure (b) over scaling factors ranging from 2x until 8x with 0.1 increments.}
\label{fig:scale_factors}
\end{figure}

\noindent \textbf{Larger scale factors}:
We further expand the scaling factors of the testing data ranging from 2x to 8x with 0.1 intervals. Figure~\ref{fig:scale_factors} illustrates that although the RMSEs for both TS and PRECT channels increase as the scale factor grows, our model consistently outperforms the bicubic interpolation up to a scale factor of 8x. 

\begin{table}[htp]
\caption{Performance comparison against standard approaches and deep learning models, trained and tested on data with random noise by 2x scaling factor. The experiment is performed over 1000 data points. The reported metric is RMSE and units are in 1E-8 for PRECT.}
\label{tab:noise_robustness}
\begin{tabular}{c|c|c}
\toprule
\textbf{Model} & \textbf{TS} & \textbf{PRECT} \\ \midrule
Bilinear & 3.4363 & 2.1406 \\
Bicubic & 4.3781 & 2.0650 \\
Base & 3.6050 & \textbf{1.3096} \\
Ours & \textbf{3.3013} & 1.3908 \\ \bottomrule
\end{tabular}
\end{table}

\subsection{Supervised Learning}
As most of the proposed machine learning downscaling models are supervised, we extend our investigation to ascertain whether the integration of channel segregation and topoclimatic attention components can improve performance in scenarios where supervised climate downscaling is feasible. 
We present preliminary results for this analysis on the CMIP6 (LR)--ERA5 (HR) data pairs. The LR data in these experiments is not a degraded version of the HR data; rather, it is a separate dataset that emulates the real data. It must be noted that while CMIP6 and ERA5 data are generally paired together for such analysis, they still do not represent a real LR -- real HR scenario due to fundamental differences between the two datasets. While CMIP6 (\url{https://esgf-node.llnl.gov/projects/cmip6/}) is a collection of various coupled climate simulation models, ERA5 \url{https://www.ecmwf.int/en/forecasts/dataset/ecmwf-reanalysis-v5} is the reanalysis data, that combines observational data and numerical predictions to create a continuous record of the past weather. For these datasets as well, we downscale TS and PRECT, while HR topography (PHIS) is the third variable that aids the model learning. 

To reduce computational complexity, we regrid ERA5 data to 1$^{\circ}$  from the original resolution of 0.25$^{\circ}$. Similarly, CMIP6 data is regridded from 1$^{\circ}$  to 3.75$^{\circ}$. Thus, the LR CMIP6 data is downscaled with a factor of 3.75x (48x96 $\rightarrow$ 128x256). Both datasets are reconciled for spatial grid and temporal scopes, and the final dataset covers daily averages from the year 1979 to 2014. Data reconciliation and re-gridding are performed with CDO, which is a collection of command line tools to manipulate and analyze climate data (\url{https://code.mpimet.mpg.de/projects/cdo/}). Moreover, to limit the extraneous perplexity, we also clip the polar regions from the data and utilize data in the region of 70$^{\circ}$ North and South of the equator. We use SRResNet as the baseline and the backbone network for all experiments~\cite{Ledig_2017_CVPR}. The model was trained on $X - X^{HR}$ pairs, where $X \in CMIP6$ and $X^{HR} \in ERA5$. In detail, the number of data is $13,140$, where the proportion of the train, validation, and test data was $0.7$, $0.1$, and $0.2$ respectively. The model was trained for 50 epochs, and the model checkpoint with minimum validation error was chosen for evaluation. 

The results in Table \ref{tab:cmip_era} show that using channel segregation and topoclimatic attention improves performance in both TS and PRECT channels. This implies that our proposed components apply to general machine learning-based climate downscaling scenarios.

\begin{table}[]
\centering
\caption{Supervised experiment on CMIP6 (LR) and ERA5 (HR) data. The components in the table are abbreviations of Channel Segregation (CS) and Topoclimatic Attention (Attention). The metric is RMSE and units are in 1E-04 for PRECT.}
\label{tab:cmip_era}

\begin{tabular}{l|c|c}
\toprule
\textbf{Model} & \textbf{TS} & \textbf{PRECT} \\ \midrule
Bilinear & 5.1150 & 3.0349 \\
Bicubic & 5.2944 & 3.0454 \\
SRResNet & 2.8484 & 2.7114 \\
SRResNet+CS & 2.7469 & 2.7156 \\
SRResNet+Att & 2.6632 & 2.7147 \\
SRResNet+CS+Att & \textbf{2.6365} & \textbf{2.7103} \\ \bottomrule

\end{tabular}
\end{table}


\begin{figure*}[t!]
\begin{minipage}[T]{0.48\textwidth}
       \centering \includegraphics[width=\linewidth]{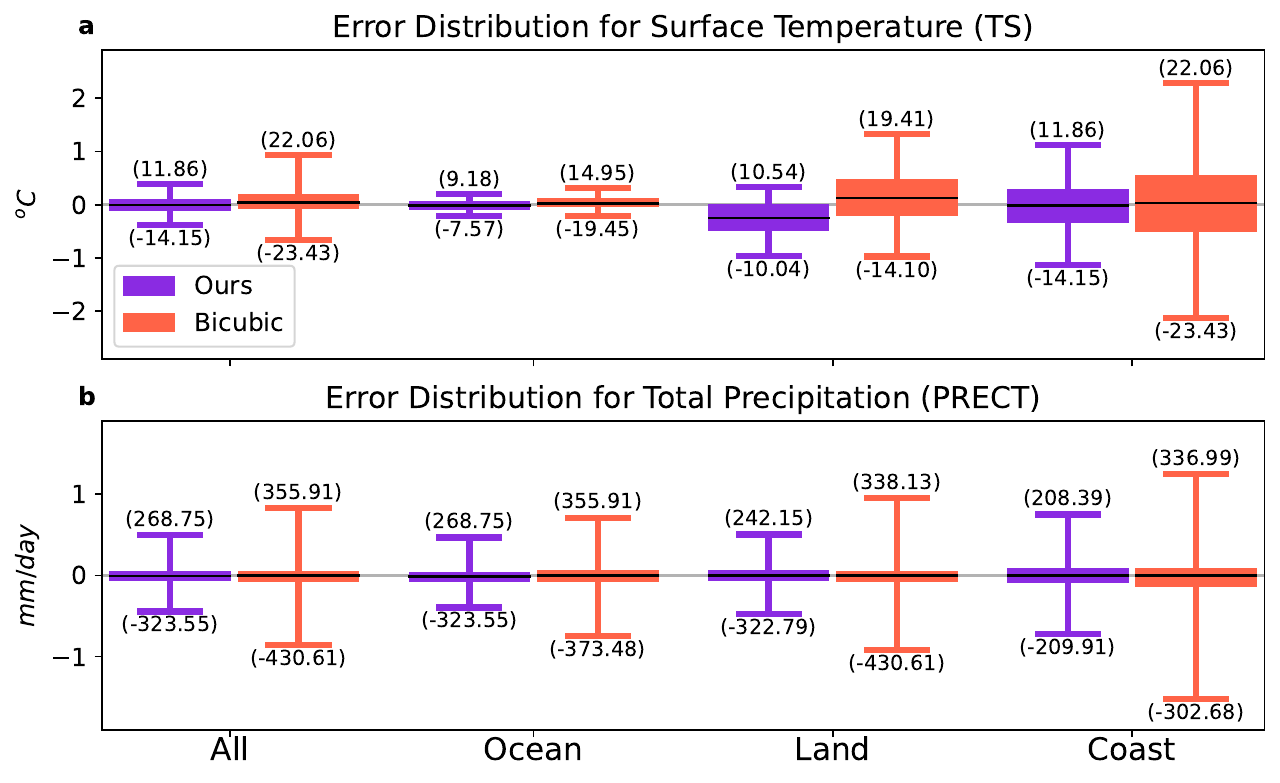}
      \caption{Comparison between the error distribution from Base (purple) and bicubic interpolation (red) for surface temperature (a) and precipitation (b). The vertical edges of solid boxes represent Q1 and Q3 of the error distribution. 
      }
      \label{error_dist}
\end{minipage}
\hspace{5mm}
\begin{minipage}[T]{0.48\textwidth}
        \centering \includegraphics[width=1.05\linewidth]{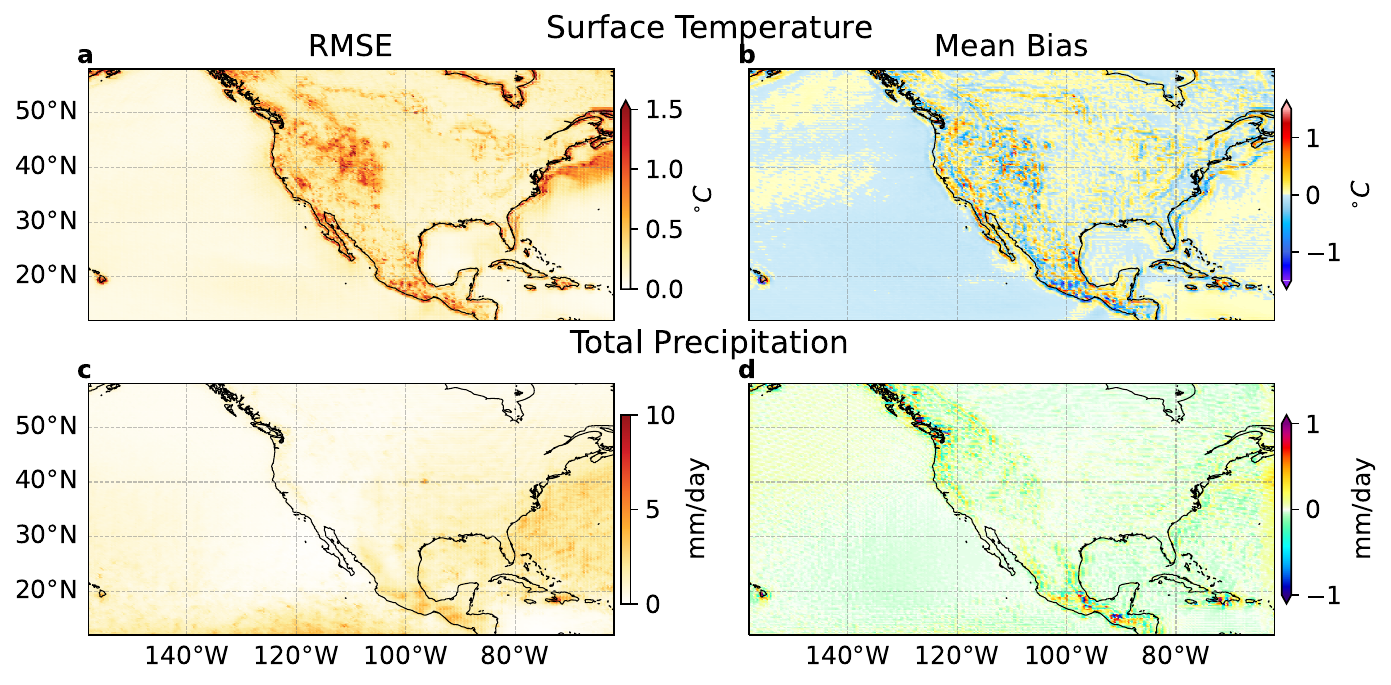}
        \caption{The spatial distribution of RMSE (a, c) and mean bias (b,d) for surface temperature (a, b) and precipitation (c, d) for downscaling by the proposed model.}
\label{error_fig}
\end{minipage}
\end{figure*}

\section{Discussion and Conclusion}

Having confirmed the overall model performance, we discuss its ability to downscale the climate variables compared to the bicubic interpolation baseline and its strengths and weaknesses in different domains within the region of interest. The main metrics used for this analysis are the RMSE and mean error (Figure \ref{error_fig}). We also look at the error distribution within domains such as oceans, land, and coastal areas Figure \ref{error_dist}. All the analyses are performed with a 2x scaling factor. The conclusions drawn from the analysis are expected to apply to other scaling factors.

\subsection{Performance in TS and PRECT channels}
\label{sec:discussion_channels}
The two variables under consideration show considerable differences in their spatiotemporal characteristics. 
As the noisier variable, the PRECT has a larger error distribution spread than TS. Our model's error distribution (purple plots) shows improvement in all error metrics (compared to the bicubic interpolation (orange plots)). The values corresponding to the $5^{th}, 25^{th}, 75^{th},$ and $95^{th}$ percentiles are lower in our model than in the bicubic interpolation, extending the confidence in our model to perform better downscaling. 

Even the maximum (value inside the parenthesis above the whisker representing the $95^{th}$ percentile) and minimum (value inside the parenthesis below the whisker representing the $5^{th}$ percentile) values of errors are smaller in our model compared those of the bicubic interpolation baseline (Figure \ref{error_dist} a,b). We emphasize that our inferences are valid for the entire region and separately in the different domains (ocean, land, and coast) within it. Both the bicubic interpolation baseline and our model demonstrate lower performance in TS over the coastal region (Figure \ref{error_dist} a). PRECT also shows similar behavior, albeit with smaller differences between different domains (Figure \ref{error_dist} b).

Over the region of interest, TS has large RMSE values over the mountainous areas of North America. The errors are also high along the warm western boundary current along the east coast (Figure \ref{error_fig} a). The former might stem from the model's inability to resolve mountainous topography well, while the latter is due to the unresolved small-scale eddies formed along the boundaries of the current system. Our model underestimates the TS over the Pacific and Atlantic Oceans (Figure \ref{error_fig} b). The RMSE spatial distribution for PRECT shows higher values at low latitudes and off the continent's eastern coast (Figure \ref{error_fig} c). The errors are introduced contrastingly on either side of the elevated terrain (\ref{error_fig} d).
The following conclusions can be drawn from the analysis of the model output:
\begin{itemize}
    \item Our model consistently outperforms bicubic interpolation in all the standard error distribution metrics.
    \item Over coastlines, both TS and PRECT show large error values. Incorporating HR coastline information into the model would improve its performance in this domain.
    \item Other hot spots for large errors are the mountainous terrains and the eddy-rich oceanic regions.
\end{itemize}

\subsection{Conclusion}
In this work, we propose a deep learning-driven climate downscaling model that does not require high-resolution ground truth data during training. Our model incorporates three components into the standard CNN model to make it adaptable to climate data: self-supervised pre-training that enables knowledge transfer from the input data and improves runtimes, channel segregation that enables learning better representations for individual weather variables, and topoclimatic attention that enforces the learning of the intricate relationship between the weather variables and topography. Extensive experiments and qualitative analysis across 2x, 3x, and 4x scaling factors show that the proposed model improves the downscaling performance by a substantial margin. 
Moreover, our preliminary analysis from the ongoing work demonstrates that the proposed components also improve the climate downscaling performance in the supervised setting. 

There are many important implications of this research as it demonstrates a high-quality downscaling ability with no dependency on ground truth data. Our work opens up possibilities such as effective downscaling of historic data, running larger climate ensembles, democratizing research by allowing organizations with limited access to computational power to build their models and promoting a greener planet by reducing energy costs. This research opens up numerous avenues for further exploration, such as mitigating the trade-off between weather variables of different natures, adding more climate variables, and incorporating temporal dimension, extending the analysis in the supervised setting, as well as improving the explainability of the deep learning model. 

\newpage
\balance
\bibliographystyle{ACM-Reference-Format}
\bibliography{main}

\appendix

\end{document}